\documentclass[aps,prb,preprint,showpacs]{revtex4}

\usepackage{graphicx} % Include figure files

\begin{document}

\title{Quantum Phase in Nanoscopic Superconductors}

\author{Zafer Gedik}
\affiliation{Faculty of Engineering and Natural Sciences, Sabanci
University, 34956 Tuzla, Istanbul, Turkey}

\date{\today}

\begin{abstract}

Using the pseudospin representation and the SU(2) phase operators
we introduce a complex parameter to characterize both infinite and
finite superconducting systems. While in the bulk limit the
parameter becomes identical to the conventional order parameter,
in the nanoscopic limit its modulus reduces to the number parity
effect parameter and its phase takes discrete values. We evaluate
the Josephson coupling energy and show that in bulk superconductor
it reproduces the conventional expression and in the nanoscopic
limit it leads to {\it quantized Josephson effect}. Finally, we
study the {\it phase flow} or {\it dual Josephson effect} in a
superconductor with fixed number of electrons.

\end{abstract}

\pacs{74.20.Fg, 74.20.-z, 74.50.+r}

\maketitle

Recent experimental works on superconducting metallic islands at
nanometer scale have established a link between bulk
superconductors and atomic nuclei as regards pairing correlations
\cite{tuominen,lafarge,ralph}. The long-range order in a bulk
superconductor can be described by an order parameter $\Delta$
which is complex and the equations have symmetry properties which
ensure that if $\Delta$ is a solution, then $e^{i\theta}\Delta$ is
also a solution \cite{bardeen, bogolubov1}. On the other hand when
size of a superconductor is reduced to nanometer scales so that
the number of the electrons is fixed, the order parameter
$\Delta=\langle c_{-{\bf k}\downarrow}c_{{\bf k}\uparrow}\rangle$
vanishes where $c_{-{\bf k}\downarrow}$ and $c_{{\bf k}\uparrow}$
are the annihilation operators for time-reversed states $\mid
-{\bf k}\downarrow\rangle$ and $\mid {\bf k}\uparrow\rangle$,
respectively. In this case, superconductivity manifests itself
with nonvanishing number parity effect parameter $\Delta_P$ where
the ground state energy of the system increases or decreases,
depending upon whether the total number becomes odd or even, by
addition of a new electron \cite{matveev,flocard}. In this work,
we propose a parameter, which unifies the order parameter $\Delta$
of the bulk limit and the number parity effect parameter
$\Delta_P$ of the nanoscopic superconductors. Introducing the
pseudospin or quasi-spin representation for the model Hamiltonian
of the theory of superconductivity \cite{anderson1,bogolubov2} and
the SU(2) phase states \cite{vourdas} we define a quantum phase
for a superconductor with discrete energy levels along with
modulus of the parameter which becomes equal to $\Delta_P$. As we
go from the nanoscopic limit to the bulk superconductor we show
that the number parity effect parameter and the SU(2) phase go to
the amplitude and the phase of the bulk order parameter,
respectively.

We are going to start with a notation which is more proper for nanoscopic
superconductors where energy levels are discrete and finite
\cite{vondelft1}. This reduced form of the BCS model was applied in
nuclear physics and it has an exact solution \cite{richardson1}.
The model Hamiltonian is
\begin{equation}
H=\sum_{j,\sigma}\epsilon_j
c_{j\sigma}^{\dagger}c_{j\sigma}
-g\sum_{j,j'}
c_{j\uparrow}^{\dagger}c_{j\downarrow}^{\dagger}
c_{j'\downarrow}c_{j'\uparrow}
\label{ham}
\end{equation}
where $g$ is the pairing coupling constant for the time-reversed states
$\mid j\uparrow\rangle$ and $\mid j\downarrow\rangle$, both having the energy
$\epsilon_j$. Here, $c_{j\sigma}^{\dagger}$ ($c_{j\sigma}$) is the creation
(annihilation) operator for state $\mid j\sigma\rangle$ where
$j\in\{1,...,\Omega\}$ and $\sigma\in\{\uparrow,\downarrow\}$.

Introducing the pseudospin variables \cite{anderson1,bogolubov2}
\begin{equation}
s_j^z=\frac{1}{2}
\left(c_{j\uparrow}^{\dagger}c_{j\uparrow}
+c_{j\downarrow}^{\dagger}c_{j\downarrow}-1\right),\;\;\;\;\;
s_j^-=c_{j\downarrow}c_{j\uparrow}=\left(s_j^+\right)^\dagger
\end{equation}
which generate the SU(2) algebra
\begin{equation}
\left[s_i^+,s_j^-\right]=2\delta_{ij}s_j^z,\;\;\;
\left[s_i^z,s_j^\pm\right]=\pm\delta_{ij}s_j^\pm,
\end{equation}
it is possible to rewrite the model Hamiltonian as
\begin{equation}
H=\sum_{j}2\epsilon_j\left(s_j^z+\frac{1}{2}\right)
-g\sum_{ij}s_i^+s_j^-.
\label{sham}
\end{equation}
We note that the mapping from the Fermi operators to the
pseudospin operators is possible as long as all single particle
states are doubly occupied. However, since the original
Hamiltonian (\ref{ham}) contains no terms which couple a singly
occupied  level to others, the only role of such states will be
blocking from pairing interaction. Therefore, the summations in
Eqn.~(\ref{sham}) are over doubly occupied or empty states. Both
the above mapping and the BCS wave function \cite{bardeen} lack
proper antisymmetrization due to separate treatment of singly
occupied states, but since the model Hamiltonian (\ref{ham}) does
not involve any scatterings into or out of such states,
antisymmetrization with respect to interlevel pair exchange and
intrapair electron exchange is sufficient.

In this work, rather than the exact solution of the problem, we are
interested in the qualitative result which has also been obtained
numerically: The ground state energy for even number
of electrons is lower in comparison to neighboring odd number
states \cite{mastellone,berger,braun,dukelsky} including degenerate
case \cite{kulik1}. Parity dependence of the condensation energy and
pairing parameters in nanoscopic superconductors was first emphasized
by von Delft {\it et al.} \cite{vondelft1} but the first correction to the
bulk limit had been obtained by Janko, Smith and Ambegaokar \cite{janko}
and Golubev and Zaikin \cite{golubev}.

Phase operators and phase states have been studied mainly in quantum optics
and possible connection of quantum phase and the mean field treatment of the
BCS Hamiltonian has been pointed out by Shumovsky \cite{shumovsky}.
Given SU(2)
algebra, for example the one generated by the components of the total spin
operator ${\bf s}=\sum_i{\bf s}_i$, we can introduce \cite{vourdas} the radial
operator defined by
\begin{equation}
s_r=\sqrt{s^+s^-}
\end{equation}
and the exponential of the phase operator given by
\begin{equation}
E=\sum_{m=-s}^{m=s}\mid S;sm+1\rangle\langle S;sm\mid.
\end{equation}
Here, $\mid S;sm\rangle$ is simultaneous eigenstate of ${\bf s}^2$
and $s_z$ operators with eigenvalues $s(s+1)$ and $m$,
respectively. In order to simplify the notation, $m$ is defined
modulo $2s+1$ so that $\mid S;ss+1\rangle=\mid S;s -s\rangle$. The
label $S$ has been introduced to distinguish them from the phase
states to be defined below. We are going to make use of the cosine
and the sine operators
\begin{equation}
Cos=\frac{1}{2}\left(E+E^{-1}\right),\;\;\;\;\;
Sin=\frac{1}{2i}\left(E-E^{-1}\right).
\end{equation}
For integer $s$ or on the so called Bose sector, the eigenstate of
$E$ with eigenvalue $\exp(-i2\pi \mu/(2s+1))$ is evaluated to be
\begin{equation}
\mid \theta;s\mu\rangle=\frac{1}{\sqrt{2s+1}}\sum_{m=-s}^{m=s}
\exp\left[i\frac{2\pi \mu}{2s+1}m\right]\mid S;sm\rangle
\end{equation}
and a similar expression holds for half integer $s$ or in the Fermi sector.

In terms of the radial and the exponential of the phase operators
for the total spin, it is possible to rewrite the interaction part
of the Hamiltonian (\ref{sham}) as $-gs_rEE^\dagger s_r$. Since
$E$ is unitary, we obviously have $EE^\dagger=I$ but our aim in
keeping $E$ and $E^\dagger$ is to define the phase properly. Now,
we introduce the superconductivity criterion as $\langle s_r
\rangle\ne 0$. We are going to prove that this definition agrees
with existing criteria for both grand canonical and canonical
superconducting systems. We are going to show that $\langle s_r
\rangle$ becomes identical to the modulus of the BCS order
parameter in the bulk limit while in the nanoscopic limit it
reduces to the number parity effect parameter $\Delta_P$ in units
of $g$. There have been several suggestions for a canonically
meaningful pairing parameter
\cite{vondelft1,braun,rossignoli,rickayzen,vondelft2,tian}. Our
definition is equivalent to that of Tian {\it et.al.} \cite{tian},
which has been proposed by Penrose and Onsager \cite{penrose} and
Yang \cite{yang} as a measure of the strength of the spontaneous
symmetry breaking field. Amico and Osterloh \cite{amico} and Zhou
{\it et.al} \cite{zhou} have calculated  the pairing correlation
function $\langle s_i^-s_j^+\rangle$ analytically by extending
Richardon's results \cite{richardson2}. We further introduce
$\langle E\rangle$  as exponential of the phase. This definition
is justified by the observation that in the grand canonical
ensemble $\langle E\rangle$ turns out to be exponential of phase
of the BCS order parameter.

We first note that $[s_r,s_z]=0$ and hence $s_r$ gives a good
quantum number even for a finite system. Secondly, $\langle s_r
\rangle$ is filling dependent even for a single, $d-$fold
degenerate level in contrast to $\Delta_P$. While $\langle s_r
\rangle=\sqrt{\nu(d-\nu+1)}$ for $\nu$ pairs, the number parity
effect parameter $\Delta_P$ is $gd/2$, independent of $\nu$. For
$\nu=d/2$, i.e. half-filling or $m=0$, the two results become
identical.

We start to our proof by examining the canonical system. In analogy to the
pairing energy in nuclear physics \cite{richardson3},
Matveev and Larkin \cite{matveev} introduced the  parity effect parameter
\begin{equation}
\Delta_P=E_{2n+1}-\frac{E_{2n}+E_{2n+2}}{2}
\end{equation}
for nanoscopic superconductors
where $E_n$ is the ground state energy for n electrons. Assuming that the
expectation value of the single particle energy part
(the first term in Eqn.~(\ref{sham})) follows a monotonic behavior so that
$T_{2n+1}\simeq(T_{2n}+T_{2n+2})/2$, the main contribution to the ground state
energy will come from the interaction part so that
\begin{equation}
\Delta_P=-g\left(\langle s_r^2\rangle_{2n+1}- \frac{\langle
s_r^2\rangle_{2n}+\langle s_r^2\rangle_{2n+2}}{2}\right).
\label{mat}
\end{equation}
Now, the eigenstates and in particular the ground state of the model
Hamiltonian will be of the form
\begin{equation}
\sum_s c_{sm}\mid S;sm\rangle \label{eig}
\end{equation}
because the interaction term commutes with ${\bf s}^2$ and $s_z$
while the single particle part commutes with the latter only and
hence $m$ is a good quantum number. We note that since $s$ is the
total spin in general it is multiply degenerate. It is possible to
calculate the expectation value of the radial operator
$s_r=\sqrt{s^+s^-}$ as
\begin{equation}
\langle s_r \rangle_n=\sum_s\mid c_{sm}\mid^2\sqrt{s(s+1)-m(m-1)}.
\end{equation}
Here, the number of electrons $n$ is a function of $m$. In BCS
theory the single particle states participating in pairing
interaction are assumed to be those in a shell of thickness
$\sim2\hbar\omega_D$, $\omega_D$ being the Debye frequency,
symmetric around the Fermi level. In this case, half of the states
are full while the half is empty and hence $m=0$. Near
half-filling where $m\simeq0$ and for $s\gg 1$, we can approximate
the square root as $s$ to give $\langle s_r
\rangle_n\simeq\sum_s\mid c_{s0}\mid^2s$. Similarly, for $s_r^2$,
with the same approximations we find that
\begin{equation}
\langle s_r^2\rangle_{2n}\simeq\langle
s_r^2\rangle_{2n+2}\simeq\sum_s \mid c_{s0}\mid^2s^2. \label{app}
\end{equation}
For $2n+1$ electrons, the mere effect of the unpaired electron is
to block one of the single-particle energy levels from pairing
which in our notation means that the corresponding spin value
becomes $s-1/2$. However, using Eqn. (\ref{mat}) this simply gives
that $\Delta_P\simeq g\langle s_r \rangle$. Therefore, the
parameter we introduced $\langle s_r \rangle$ (multiplied by the
pairing coupling constant $g$) is identical to the number parity
effect in the proper limit.

Next, we examine the grand canonical case or the thermodynamic
limit. In its present form the model Hamiltonian (\ref{sham})
commutes with $s_z$ and therefore $m$ is a good quantum number or
equivalently the number of electrons is a conserved quantity. To
make a connection with the BCS order parameter we are going to
replace the interaction part of the Hamiltonian by
\begin{equation}
-g(s_rE\Delta^*+\Delta E^\dagger s_r-\mid\Delta\mid^2)
\label{tham}
\end{equation}
which is nothing but the standard mean field approximation since
$s_rE=s^+$ and $E^+s_r=s^-$. The BCS wave function describes a
state with totally indefinite number of particles but with a
definite phase. We can project the BCS states onto states of
definite particle number by taking the Fourier transform with
respect to the phase \cite{anderson2} and that is why particle
number $N$ and phase $\phi$ are conjugate variables with an
uncertainty relation $\delta N\delta\phi\simeq 1$. It has been
shown that in the thermodynamic limit the ground state of the BCS
Hamiltonian (\ref{sham}) is also the ground state of the mean
field Hamiltonian whose interaction part is given by Eqn.
(\ref{tham}) \cite{thirring}. These are nothing but the phase
states which we have defined above. In our case this result can be
verified by observing that near half-filling and at large $s$, we
have $\left[s_r,E\right]\simeq0$. Therefore, we evaluate the
expectation value of $s_rE$ in state $\mid\theta;s\mu\rangle$ and
find that
\begin{equation}
\frac{\exp(-i2\pi \mu/(2s+1))}{2s+1}\sum_m \sqrt{s(s+1)-m(m-1)}.
\end{equation}
We identify the phase $-2\pi \mu/(2s+1)$ as $\phi$ and the factor
in front (the sum divided by $2s+1$) as the modulus of the order
parameter $\mid\Delta\mid$. This completes our argument on the
relation of $\langle s_r\rangle$ and $\langle E\rangle$ to
$\Delta_P$ and $\Delta$ except one point:
What happens to $\langle E\rangle$ for a system with
discrete energy levels but yet with indefinite number of
electrons? We note that this not the thermodynamic limit. The
system is finite but yet the number of electrons is not fixed.
Such a situation can be realized through a Josephson junction.

The origin of the Josephson interaction is single-particle tunneling
electron pairs. At low energies, single-particle tunneling interaction
lead to two contributions both of which are second order processes. The
first one, where an electron goes from one superconductor to the other and
returns, leads to proximity effect. The second one is the Josephson tunneling of
two electrons from one superconductor to the other.
The only effect of the first process is to renormalize
the single-particle energies. Furthermore, there is no net current associated
with this process. We can evaluate the explicit contributions of these two
processes by considering two superconductors, both of which are described by
the model Hamiltonian (\ref{sham}) so that we are going to denote the total
Hamiltonian as $H_0$. Let us consider a tunneling interaction of the form
\begin{equation}
V=t\sum_{j,j',\sigma}(c_{2j'\sigma}^\dagger c_{1j\sigma}+
c_{1j\sigma}^\dagger c_{2j'\sigma})
\end{equation}
where $c_{1j\sigma}(c_{2j\sigma})$ is the annihilation operator
for state $\mid j\sigma\rangle$ of the first (second)
superconductor. One way to introduce $V$ perturbatively is to use
the unitary transformation method \cite{wagner} where the second
order Hamiltonian takes the form $H_0+[V,\Omega]/2$. Here, the
anti-Hermitian operator $\Omega$ is given by
\begin{equation}
\Omega=\sum_{m_1,m_2,n_1,n_2}\frac{ \mid m_1m_2\rangle\langle
m_1m_2\mid V \mid n_1n_2\rangle\langle n_1n_2\mid}
{\epsilon_{m_1m_2}^{(0)}-\epsilon_{n_1n_2}^{(0)}}
\end{equation}
where $\mid n_1n_2\rangle$ denotes the ground state of $H_0$ with
$n_1$ electrons in the first superconductor and $n_2$ electrons in
the second and $\epsilon_{n_1n_2}^{(0)}$ is the corresponding
energy eigenvalue of the combined system. Since we are interested
in low energy excitations, at each step we project the system into
its ground state. The two contributions we discussed above, the
proximity and Josephson processes, can easily be calculated.
Repeating the approximations we did in Eqn. (\ref{app}), we find
that the strength of both terms are given by
$-t^2/(\Delta_{1P}+\Delta_{2P})=\varepsilon_J$. In particular the
Josephson interaction term can then be written as
$\varepsilon_J(E_1E_2^{-1}+E_1^{-1}E_2))/2$ where $E_i$ is the
exponential of the phase operator in the $i^{th}$ superconductor.
We immediately observe that for phase state
$\mid\phi_1\phi_2\rangle$, expectation value of this term is
simply $\varepsilon_J\cos(\phi_1-\phi_2)$. To simplify our final
analysis let us assume that one of the superconductors is large so
that it can be described by the BCS state with a fixed phase
$\phi$ which we can assume to be zero without loss of generality.
Then for the other we can write down an effective Hamiltonian
\begin{equation}
H_{eff}=\sum_j2\tilde{\epsilon}_j\left(s_j^z+\frac{1}{2}\right)
-gs^+s^-+4\varepsilon_C\left(s_z-\langle
s_z\rangle\right)^2+\varepsilon_JCos
\end{equation}
where $\varepsilon_C$ is single-electron charging energy of the
island and $\tilde{\epsilon}_j$ is renormalized single-particle
energy. The Josephson current $I_J=2e\langle\dot{s}_z\rangle$ can
be easily calculated as $I_J=2e\varepsilon_J\langle
Sin\rangle/\hbar$ where $Sin$ is the sine operator. Therefore, in
the bulk limit where eigenstates are nearly phase states, we
recover the conventional expression for the Josephson current
\cite{josephson1,josephson2}. Eigenstates of the Hamiltonian
composed of the first three terms of $H_{eff}$ are still given by
(\ref{eig}) where $m$ is a good quantum number. Hence, $H_{eff}$
is nothing but tight-binding Hamiltonian with nearest neighbor
$(m\pm1)$ hopping matrix element $\varepsilon_J/2$. The nature of
the eigenstates depends upon the on-site energies. For example,
for quadratic dependence of energy eigenvalues (in the absence of
$\varepsilon_JCos$ term) on $m$, which would be the case for flat
$\tilde{\epsilon}_j$, we can find the exact eigenvalues and
eigenstates of $H_{eff}$. In this case we obtain a tight-binding
Hamiltonian with on-site energies having quadratic dependence on
site index and we can find the solution by observing that the
expansion coefficients of the Mathieu function $ce_{2n}$ satisfy a
recursion relation which is identical to the characteristic
equation of $H_{eff}$ \cite{grad}.

It is clear that $I_J$ vanishes for any state which can be written
as a linear combination of $\mid S;sm\rangle$ states with real
expansion coefficients. These are nothing but bound states in
$S-$space. On the other hand for propagating states, like $\mid
\theta;s\mu\rangle$, $I_J$ is non-zero. For small enough systems a
very interesting situation may arise because discreteness of $\mu$
and hence quantized $I_J$ values might be observed. In other
words, if the number of the single particle energy levels and
hence $s$ is not too big, we can measure a {\em quantized
Josephson current}. A single electron transistor with a small
enough superconducting island can be used to see quantization
effect. Another possibility is to measure the Josephson plasma
oscillations between a bulk and nanoscopic superconductor
\cite{kulik2,barone}. Recent first principle calculations for
structural and electronic properties of aluminum covered single
wall carbon nanotubes show that a stable metallic ring can be
formed \cite{bagci}. These structures can also allow us to observe
effect of phase quantization. Coulomb interaction works in the
direction to suppress the current but using an external electric
field relative strength of the Josephson interaction can be
increased. One possibility is to measure the Josephson current
through one dimensional array of aluminum rings formed around a
carbon nanotube.  In general, any physical quantity depending upon
the phase is a candidate to observe quantization. For BCS gap
$\Delta=2\hbar\omega_De^{-1/\lambda}$ and level spacing $\delta$
satisfying $\delta\lesssim\Delta$, assuming for example that we
can resolve discreteness of the phase angle for
$2\hbar\omega_D/\delta\simeq 1000$ states in the Debye shell at
the Fermi level, we evaluate $\lambda$ to be $\gtrsim 0.14$. For
larger $\lambda$, we can go to smaller sizes or less number of
states and hence there is more chance to observe quantization
effects.

If $N$ and $\phi$ are conjugate variables and the Josephson effect
is a phenomenon relevant to fixed $\phi$ and indefinite $N$, what
is its dual effect where $N$ is fixed but $\phi$ is indefinite?
When a nanoscopic superconductor is coupled by Coulomb interaction
to another superconductor, there appears a second order effect
which is analogue or {\it dual of Josephson effect} where particle
numbers are fixed but the phases are not determined. To make the
analogy complete let us consider an interaction term of the form
$\varepsilon_D(F+F^{-1})$ where $\varepsilon_D$ is the dual
Josephson interaction energy and $F$ is dual to the operator $E$
and it is defined by
\begin{equation}
F=\exp\left[i\frac{2\pi}{2s+1}s_z\right].
\end{equation}
It is easy to show that
$F\mid\theta;s\mu\rangle=\mid\theta;s\mu+1\rangle$ \cite{vourdas}.
We can evaluate the {\it phase current}
$\langle\dot{\theta}_z\rangle$ where
\begin{equation}
\theta_z=\sum_{\mu=-s}^{s}\mu\mid\theta;s\mu\rangle\langle\theta;s\mu\mid,
\end{equation}
in complete analogy to Josephson current as
$i\varepsilon_D(F^{-1}-F)/\hbar$. This interaction is similar to
the van der Waals force between two molecules which is a
manifestation of discreteness of electronic energy levels. In the
superconducting state, intragrain single particle excitation
spectra are modified due to the number parity effect and hence
there appears an additional interaction due to pairing. In other
words, dual Josephson effect refers the attractive interaction
between two superconductors due to virtual Cooper pair breaking
(as a result of Coulomb interaction between the superconductors)
where interaction energy $\epsilon_D$ is of the order of the ratio
of Coulomb interaction squared to superconducting gap or number
parity effect parameter. This effect might also have relevance to
atomic nuclei when they approach close enough so that Coulomb
force becomes appreciable.

In conclusion, we proposed a complex parameter to describe pairing
correlations in a fermionic system. We showed that our definition
agrees with the existing parameters in the canonical and grand
canonical descriptions. We predicted possible quantization in
Josephson effect in the nanoscopic limit. We further analyzed the
dual Josephson effect a nanoscopic superconductor and interpreted
the resulting expression in terms of quantum phase flow. Recently,
the complex parameter introduced this work has been used to study
quantum entanglement a paired finite Fermi system \cite{gedik}.

\begin{acknowledgments}
The author thanks to A. Baratoff, I.O. Kulik, and A.S. Shumovsky
for helpful discussions and acknowledges the constant support by
B. Bilgin and O. G{\" u}lseren. This work has been supported by
the Turkish Academy of Sciences, in the framework of the Young
Scientist Award Program (MZG/T{\"U}BA-GEB{\.I}P/2001-2-9).
\end{acknowledgments}


\begin{thebibliography}{99}

\bibitem{tuominen} M.T. Tuominen, J.M. Hergenrother, T.S. Tighe, and M.
Tinkham, Phys. Rev. Lett. {\bf 69}, 1997 (1992).
\bibitem{lafarge} P. Lafarge, P. Joyez, D. Esteve, C. Urbina, and M.H. Devoret,
Phys. Rev. Lett. {\bf 70}, 994 (1993).
\bibitem{ralph} D.C. Ralph, C.T. Black, and M. Tinkham, Phys. Rev. Lett.
{\bf 74}, 3241 (1995); {\bf 76}, 688 (1996); {\bf 78} 4087 (1997).
\bibitem{bardeen} J. Bardeen, L.N. Cooper, and J.R. Schrieffer, Phys. Rev.
{\bf 108}, 1175 (1957).
\bibitem{bogolubov1} N.N. Bogolubov, Sov. Phys. JETP {\bf 7}, 41 (1958).
\bibitem{matveev} K.A. Matveev and A.I. Larkin, Phys. Rev. Lett.
{\bf 78}, 3749 (1997).
\bibitem{flocard} H. Flocard, {\it Atomic Clusters and Nanoparticles,
Les Houches Lectures Session  LXXIII}, edited by C. Guet, P. Hobza,
F. Spiegelman, and F. David (Springer, Berlin, 2001), p. 221.
\bibitem{anderson1} P.W. Anderson, Phys. Rev. {\bf 110}, 985 (1958).
\bibitem{bogolubov2} N.N. Bogolubov, Sov. Phys. JETP {\bf 34}, 73 (1958).
\bibitem{vourdas} A. Vourdas, Phys. Rev. A {\bf 41}, 1653 (1990).
\bibitem{vondelft1} J. von Delft, A.D. Zaikin, D.S. Golubev, and W. Tichy,
Phys. Rev. Lett. {\bf 77}, 3189 (1996).
\bibitem{richardson1} R.W. Richardson, Phys. Lett. {\bf 3}, 277 (1963).
\bibitem{mastellone} A. Mastellone, G. Falci and R. Fazio, Phys. Rev. Lett.
{\bf 80}, 4542 (1998).
\bibitem{berger} S.D. Berger and B.I. Halperin, Phys. Rev. B {\bf 58}, 5213
(1998).
\bibitem{braun} F. Braun and J. von Delft, Phys. Rev. Lett. {\bf 81}, 4712
(1998).
\bibitem{dukelsky} J. Dukelsky and G. Sierra, Phys. Rev. Lett. {\bf 83},
172 (1999).
\bibitem{kulik1} I.O. Kulik, H. Boyaci, and Z. Gedik, Physica C {\bf 352}, 46
(2001).
\bibitem{janko} B. Jank{\'o}, A. Smith, and V. Ambegaokar, Phys. Rev. B
{\bf 50}, 1152 (1994).
\bibitem{golubev} D.S. Golubev and A.D. Zaikin, Phys. Lett. A {\bf 195}, 380
(1994).
\bibitem{shumovsky} A.S. Shumovsky, {\it Modern Nonlinear Optics,
Advances in Chemical Physics}, edited by M.W. Evans ( Wiley, 2001),
2nd ed., Vol. 119, Part 1, p. 395.
\bibitem{rossignoli} R. Rossignoli, N. Canosa, and P. Ring, Ann. Phys. (N.Y.)
{\bf 275}, 1 (1999).
\bibitem{rickayzen} G. Rickayzen, {\it Theory of Superconductivty}
(Interscience, New York, 1965).
\bibitem{vondelft2} For a review, see J. von Delft, Ann. Phys. (Leipzig)
{\bf 10}, 219 (2001).
\bibitem{tian} G.-S. Tian, L.-H. Tang, and Q.-H. Chen, {\it Europhys. Lett.}
{\bf 50}, 361 (2000)
\bibitem{penrose} O. Penrose and L. Onsager, {\it Phys. Rev.} {\bf 104},
576 (1956).
\bibitem{yang} C.N. Yang, {\it Rev. Mod. Phys.} {\bf 34}, 694 (1962).
\bibitem{amico} L. Amico and A. Osterloh, Phys. Rev. Lett. {\bf 88}, 127003
(2002).
\bibitem{zhou} H.-Q. Zhou, J. Links, R.H. McKenzie, and M.D. Gould,
{Phys. Rev. B} {\bf 65}, 060502(R), 2002.
\bibitem{richardson2} R.W. Richardson, J. Math. Phys. (N.Y.) {\bf 6}, 1034
(1965).
\bibitem{richardson3} R.W. Richardson, Phys. Rev. Lett. {\bf 14}, 325 (1965).
\bibitem{anderson2} P.W. Anderson, Phys. Rev. {\bf 112}, 1900 (1958).
\bibitem{thirring} W. Thirring, Comm. Math. Phys. {\bf 7}, 181 (1968).
\bibitem{wagner} M. Wagner. {\it Unitary Transformations in Solid State Physics}
(North Holland, Amsterdam, 1986), p.11.
\bibitem{josephson1} B.D. Josephson, Phys. Rev. Lett. {\bf 1}, 251 (1962).
\bibitem{josephson2} B.D. Josephson, {\it Superconductivity}, edited by
R.D. Parks (Dekker, New York, 1969), p.423.
\bibitem{grad} I.S. Gradshteyn and I.M. Ryzhik, {\it Table of integrals,
Series, and Products} (Academic Press, San Diego, 1980), p. 992.
\bibitem{kulik2} I.O. Kulik and I.K. Yanson, {\it The Josephson Effect in
Superconductive Tunneling Junctions}, translated by P. Gluck
(Israel Program for Scientific Translations,
Jerusalem, 1972).
\bibitem{barone} A. Barone and G. Paterno, {\it Physics and
Applications of the Josephson Effect} (John Wiley, 1982).
\bibitem{bagci} V.M.K. Bagci, O. G{\"u}lseren, T. Yildirim, Z.
Gedik, and S. Ciraci, Phys. Rev. B {\bf 66}, 045409 (2002).
\bibitem{gedik} Z. Gedik, Solid State Commun. {\bf 124}, 473
(2002).

\end{thebibliography}
\end{document}